\documentstyle[aas2pp4]{article}

\slugcomment{Submitted to The Astrophysical Journal}

\lefthead{LEVINE, RAPPAPORT, \& ZOJCHESKI}
\righthead{Orbital Decay in LMC X-4}

\def\RXTE{{\it RXTE}}
\def\EXOSAT{{\it EXOSAT}}

\def\Ginga{{\it Ginga}}
\def\Rosat{{\it ROSAT}}

\begin{document}

\title{Orbital Decay in LMC X-4}
\author{Alan M. Levine, Saul A. Rappaport, \& Goce Zojcheski}

\affil{
Department of Physics and Center for Space Research,
Massachusetts Institute of Technology, Cambridge, MA 02139; aml@space.mit.edu
} 

\authoraddr{Center for Space Research, M.I.T., Room 37-575,
   Cambridge, MA 02139-4307}
\authoremail{aml@space.mit.edu}

\begin{abstract}

We report on the results of observations of the binary X-ray pulsar
LMC~X-4 with the {\it Rossi X-ray Timing Explorer}.  Our analysis of
the Doppler delays of the 13.5-s X-ray pulsations yields the most
accurate determination of the LMC~X-4 orbital parameters available to
date.  The epoch of orbital phase zero for the 1.4 day orbit is
determined with an uncertainty of $\sim 20$ s, and is combined with 5
earlier determinations of the epoch of phase zero to obtain the rate
of change in the orbital period: $\dot{P}_{orb}/P_{orb} = (-9.8\pm
0.7) \times 10^{-7}$ yr$^{-1}$.  This is the first high significance
measurement of the rate of change of the orbital period in LMC~X-4.

We present data on one of three strong X-ray flares as well as
energy-dependent pulse profiles for both non-flaring and flaring time
intervals.  The pulse profiles during the non-flaring time intervals
are typically strikingly different from the flare profiles, but at
other times can be similar.  Possible origins of the flaring behavior
are discussed.

We reconsider the orbital decay of LMC X-4 in the context of tidal
evolution.  We find that, while the orbital decay is most likely
driven by tidal interactions, the asynchronism between the orbit and
the rotation of the companion star is probably maintained by the
evolutionary expansion of the companion star, just as is thought to be
the case for Cen~X-3 and SMC~X-1.  For LMC~X-4, we find that the
evidence favors the companion star being in a late stage of its evolution
on the main sequence.

The orbital decay timescale of about 1,000,000 years implies that the
Roche lobe will move catastrophically deep into the atmosphere of the
companion within a few times $10^4$ years.  This short X-ray lifetime
must be similar to the lifetimes of SMC~X-1 and Cen~X-3 which have
decay timescales of 300,000 and 550,000 years, respectively, and may
be typical of all massive X-ray binaries in Roche-lobe or near
Roche-lobe contact.

\end{abstract}

\keywords{Stars:individual(LMC~X-4) --- stars:neutron --- X-rays:stars}

\section{INTRODUCTION}

The history of the orbital period of a binary star system can tell
much about the physics of the stellar components and their mutual
interactions.  In most binaries, the evolution of the orbital period
is too slow to be detectable, but there are systems in which this
evolution can be measured.  Among X-ray binaries, this evolution is
apparent in a number of cases.  Such binaries include Her X-1, which
has an intermediate mass companion to the compact star, and Cyg X-3,
which has a companion whose mass is not well determined, although it
may be a high mass Wolf-Rayet star (\cite{dee91,kit95}; \cite{vke96}
and references therein).  In systems such as Cen X-3 and SMC X-1,
which comprise X-ray pulsars orbiting around high mass companions,
extraordinary sensitivity to orbital changes and short orbital
evolutionary timescales combine to allow orbital period changes to be
measured with high precision.

Indeed, highly significant detections of orbital period changes have
been made in Cen~X-3 and SMC~X-1.  Kelley et al. (1983b) determined
that the orbital period, $P_{orb}$, of Cen~X-3 is changing at a rate
$\dot{P}_{orb}/P_{orb} = (-1.8\pm 0.08)\times 10^{-6}$ yr$^{-1}$,
while Levine et al. (1993) established that $\dot{P}_{orb}/P_{orb} =
(-3.36\pm 0.02)\times 10^{-6}$ yr$^{-1}$ for SMC~X-1.  Attempts to
measure $\dot{P}_{orb}/P_{orb}$ for LMC~X-4 and 4U~1538-52 have
resulted in only marginal detections or upper limits (see discussion
in \cite{woo96}; \cite{saf96,cor93}).

Levine et al. (1993) concluded that the rapidly decreasing orbital
period in SMC X-1 was most likely caused by tidal interactions between
the neutron star and its massive companion.  Tidal evolution requires
asynchronism between the orbital motion and the rotation of the
companion star.  The nuclear evolution of the companion in its
hydrogen shell burning phase drives rapid expansion of the star and
hence a rapid increase in the stellar moment of inertia.  Levine et
al. (1993) argued that this results in a decrease in the rate of
rotation of the companion star so that it would be rotating more
slowly than required for synchronism with the orbital motion, and
would set the conditions for orbital decay by tidal torques.  The
Darwin instability was not needed to explain the orbital decay.  The
inference that hydrogen burning is taking place in a shell within Sk
160, the companion to SMC X-1, was reinforced by evidence that its
radius is likely to be larger than that of a main-sequence star of
mass within the range of estimates for Sk 160.

Tidal interactions most likely also drive the orbital decay of Cen X-3
(\cite{kel86,lev93}). However, Levine et al. (1993) noted that the $12
\pm 2$ R$_{\sun}$ radius of the companion star in this system is
smaller than the $17 \pm 1$ R$_{\sun}$ radius of Sk 160, even though
both stars have similar masses, and that this smaller radius was more
likely to be consistent with the radius of a $\sim20$ M$_{\sun}$ star
late in its core hydrogen-burning phase. Other differences between the
SMC~X-1 and Cen~X-3 systems were also noted, which generally made it
difficult to reach firm conclusions about the exact evolutionary state
of the Cen~X-3 system.

Safi-Harb, \"{O}gelman, \& Dennerl (1996) discussed orbital decay of
LMC~X-4 in the context of conservative mass transfer and tidal
evolution and made a comparison between LMC~X-4 and Cen~X-3.  Woo et
al. (1996) noted that their estimate for the orbital period derivative
of LMC~X-4, i.e., $\dot{P}_{orb}/P_{orb} = (-5.3\pm 2.7) \times
10^{-7}$ yr$^{-1}$, could not exclude a fairly small value for
$\dot{P}_{orb}/P_{orb}$.  Woo et al. also noted that the estimated
radius of the companion of LMC X-4, R $\sim 8$ R$_{\sun}$, was
significantly smaller than that of Sk 160 even though the masses are
not much different, and was in the range expected for a $\sim 15$
M$_{\sun}$ star in the late stages of its life on the main sequence.
Woo et al. concluded that the orbital decay rate of LMC~X-4 was
consistent with the companion star being in the late main-sequence
phase.

LMC X-4 is similar in many respects to Cen X-3 and SMC X-1.  It is a
highly luminous X-ray source in the Large Magellanic Cloud
(\cite{gia72}) whose optical counterpart is a 14th magnitude OB star
(\cite{pak76,pes76}).  Pulsations of the X-ray emission with a 13.5 s
period demonstrate that the compact object is a neutron star
(\cite{kel83a}).  Optical photometric and spectroscopic variations
which are periodic and X-rays eclipses which recur with the same
period of 1.408 days revealed the binary nature and orbital period of
the system (\cite{che77,li78,whi78}).  A 30.5-day periodicity in the
X-ray intensity, discovered by Lang et al. (1981), is most likely due
to a precessing tilted accretion disk which periodically blocks our
line of sight to the neutron star.

Observations of LMC X-4 which we have obtained using the {\it Rossi
X-ray Timing Explorer} (\RXTE\,) are described in Section 2 of this
paper.  Analyses and results pertaining to the timing data, a large
flare, and pulse profiles are presented in Section 3.  The results are
discussed in Section 4.

\section{OBSERVATIONS}

The \RXTE\ All-Sky Monitor light curve (\cite{lev96}) of LMC X-4 was
used to predict the times of the LMC~X-4 30.5-d cycle high states.  A
16-day interval, 1998 October 14 to 1998 October 30, was selected so
as to be centered on the time of maximum intensity of one of the high
states.  LMC X-4 was then observed with the Proportional Counter Array
(PCA; \cite{jah96}) and High-Energy X-ray Timing Experiment (HEXTE;
\cite{rot98}) on \RXTE\ during 47 intervals within this 16-day time
interval.  All five Proportional Counter Units (PCUs) were operating
during the great majority of the observations, although a few
observations were carried out with only 3 or 4 operational PCUs.

The results presented here were obtained from our analysis of the PCA
data.  These data were telemetered in ``Goodxenon1'' and
``Goodxenon2'' modes which preserve the 1 microsecond time resolution
and the inherent energy resolution of the detectors.

The average count rate attributed to LMC X-4 for each observation
outside of eclipse is shown in Figure 1 for each of two energy bands.
To make this plot, background was subtracted from the observed count
rates using the faint source version (L7\_240) of the program
``pcabackest.''  The source intensity varied over the 16 day time
interval in a manner more or less consistent with the general pattern
of the 30.5~d cycle, i.e, an approximately triangular waveform (see
\cite{lan81}).  The counting rates attributable to LMC~X-4 reached
peak values of $\sim 100$ and $\sim 180$ counts s$^{-1}$ in the 2 - 8
and 8 - 20 keV energy bands, respectively, near the expected time of
maximum intensity.  We note that the similarity of the variation in
the two energy bands indicates that the variation cannot be attributed
solely to photoelectric absorption by neutral matter of cosmic element
abundances.

We estimated the X-ray luminosity of LMC X-4 by integrating a model
spectrum consisting of an exponentially cut off power law and a
Gaussian line at $\sim6.4$ keV derived from a fit to the observed
spectrum corrected for background.  Inclusion of low energy absorption
in the spectral model generally resulted in a negligibly low value for
the absorbing column-density parameter.  The result for the average
spectrum obtained from the observations of 1998 October 21 (Julian
Date 2451107.85; \RXTE\ observation no. 30085-01-15-00) is $L \sim 2.3
\times 10^{38}$ ergs s$^{-1}$ (2 - 25 keV) for an assumed distance of
50 kpc.

\section{ANALYSIS AND RESULTS}

\subsection{Timing Analysis}

Pulse timing analyses have been performed with
non-background-subtracted data.  The ``Goodxenon'' events were binned
into 1/16 s time bins for two energy bands, 2 - 8 keV and 8 - 20 keV.
Only events from the front xenon-filled layer of each PCU (i.e., L1
and R1) were used for the 2 - 8 keV band, while events from all three
xenon-filled layers were used for the 8 - 20 keV band.

For the pulse-timing analysis, the time of each bin of data was
corrected for the time delay incurred by the observation of the
pulsations from LMC~X-4 at the spacecraft position rather than at the
barycenter of the solar system.  In making these corrections we used
the coordinates of LMC~X-4 measured by Bradt, Doxsey, \& Jernigan
(1979) converted to the equinox J2000.0 reference frame, viz.,
$\alpha_{J2000} = 5^{\rm h}\ 32^{\rm m}\ 49\fs2$, $\delta_{J2000} =
-66\arcdeg\ 22\arcmin\ 15\arcsec$.

Subsequent stages of the pulse timing analysis were performed multiple
times as part of an iterative process in which the orbital and pulse
parameters were gradually refined until, as noted below, the changes
in the parameter values were negligible.  Each iteration of the
procedure began with the further correction of the observation times
for the orbital motion of the pulsar according to a provisional
orbital ephemeris for LMC~X-4.  This yields times that would have been
recorded by an instrument moving with an approximately constant
velocity relative to the pulsar.  The corrected times were then used
to fold subsets of the data according to a provisional pulsar rotation
frequency and frequency derivatives up to second order.

The count rate data from each of the 47 observations were folded to
produce pulse profiles. Data taken while the source was in eclipse or
when a strong flare was occurring were not included in the profiles.
Smearing of these profiles due to inaccuracies in the final values of
the parameters was less than 0.1 pulse periods.

A pulse template (see Figure 2) was somewhat arbitrarily constructed
by averaging the profiles from the first 12 observations (interval A
in Figure 3) which yielded high quality phase measurements.  The
template was then cross correlated with the individual profiles, and a
precise value for the phase corresponding to the peak of the cross
correlation function was obtained via quadratic interpolation of the
bin with the peak value and the nearest neighbor bins.  We neglected
corrections for the zero-delay autocorrelation peak that should be
present in the cross correlation of the template with any of the 12
profiles used to form the template, since we believe that such
corrections are presently unimportant.

The phases of the cross correlation function peaks should be regarded
as differences between the observed pulse phases and those of a model
pulsar in a model orbit.  Phase differences can result from
differences between the actual orbit and the model orbit; such phase
differences can properly be regarded (after being multiplied by the
pulse period) as pulse arrival time differences.  Phase differences
which result when the spin frequency of the actual pulsar differs from
that of the model pulsar can properly be regarded as pulsar rotation
phase differences.  There can also be other causes of phase
differences, such as changes in the intrinsic beam pattern of the
pulsar or simply statistical fluctuations in the observed count rates.

We performed the pulse timing analysis on the data from each of the
two energy bands separately.  The cross correlation phase differences
with respect to a best-fit model described below are shown in Figure
3.  This figure clearly shows that phase differences which are small
in magnitude, i.e., $< 0.1$ cycles, and relatively constant from one
observation to the next were obtained for a majority but not for all
of the observations.  The large scatter of the phase differences for
the first five observations, which were obtained when the overall
source intensity was low, cannot be entirely attributed to lack of
sufficient signal.  Rather, the pulse profile was significantly
different from the template profile during three of the five
observations, while the pulsations were very weak during two of the
observations.

The pulse phases obtained from the last $\sim3$ days of the
observations are not consistent between the results for the two energy
bands and also show both significant changes (relative to our best fit
model) as well as sudden jumps.  These problems are mostly caused by
substantial changes in the pulse profile which are apparent by direct
inspection and by examination of the relative strengths of the
harmonics of the pulse frequency in Fourier transforms.  The pulse
profile changes are discussed further in Section 3.3.

We fit the results for the 28 observation intervals in which the phase
differences appeared to be well-behaved (solid black circles in Figure
3) with a general circular orbit model to obtain corrections to the
provisional orbital and pulse period ephemerides.  The model included
parameters for the epoch of orbital phase zero, $T_{\pi/2}$, and
projected orbital radius, $a_{x} \sin i$, as well as parameters for the
pulse epoch, frequency, and two frequency derivatives.  Corrections to
model parameters were obtained using a linearized (with respect to a
provisional model) unweighted least-squares fitting procedure.  The
root mean square deviation of the observed arrival times from the
model arrival times was used as an estimate of the observational
uncertainty in each arrival time.  The latter uncertainties were then
propagated through the linear least-squares procedure to estimate $1
\sigma$ errors in the orbital and spin frequency parameters.  The
folding and fitting procedures were iterated using corrected
parameters until the corrections became small in comparison with their
estimated uncertainties.

The parameters for the best fit circular orbit model are given in
Table 1.  We derive a value for $a_{x} \sin i$ of $26.34 \pm 0.02 (1
\sigma)$ lt-s which is consistent with the value of $26.31 \pm 0.07$
lt-s derived from \Ginga\ observations by Levine et al. (1991), with
the value of $26.27 \pm 0.04$ lt-s derived from an earlier \Ginga\
observation by Woo et al. (1996), and with slightly less precise
values derived from \Rosat\ data by Safi-Harb, \"{O}gelman, \& Dennerl
(1996) and Woo et al. (1996).  The weighted average of the fitted
values of the epoch of orbital phase zero for the two energy bands is
$T_{\pi/2} = {\rm JD}\ 2,451,111.86579\pm 0.00010$ (TT).  The
estimated error in $T_{\pi/2}$ is calculated assuming statistical
independence of the errors in the results from each energy band.
These errors are, in turn, estimated on the basis of the scatter of
the observed arrival times relative to the model arrival times as
described above.  However, the difference between the values of
$T_{\pi/2}$ obtained for the two energy bands indicates that the
magnitude of the error of the weighted average may be underestimated
by a factor as large as $\sim 2$, and we therefore use an error
estimate of 0.00020 d ($= 17$ s)in our determination of the orbital
decay (see below).  In any case, this result is significantly more
accurate than previous determinations of the epoch of phase zero of
LMC~X-4.  \placetable{table1}

We have also fit the pulse timing results from the 28 observations
with a model that allows a small eccentricity $e$ by including
parameters that are proportional to $e \sin \Omega$ and $e \cos
\Omega$ as coefficients of $\cos 2\omega_{orb}t$ and $\sin
2\omega_{orb}t$, respectively.  Here, $\Omega$ is the longitude of
periastron and $\omega_{orb}$ is the orbital angular frequency.  No
significant eccentricity was found, so that we report an upper limit
$e < 0.003$ ($2 \sigma$).

The resultant Doppler delay data, with the delays due to the best-fit
circular orbit added back in, are plotted in Figure 4.

Our determination of the epoch of orbital phase zero is shown together
with previously determined epochs of orbital phase zero for LMC X-4 in
Figure 5.  The solid curve is the best-fit quadratic function of time
(see Table 1).  The coefficient of the quadratic term of the fitting
function implies $\dot{P}_{orb}/P_{orb} = (-9.8\pm 0.7) \times
10^{-7}$ yr$^{-1}$.  This represents the first highly significant
evidence for orbital decay in LMC X-4.  It is consistent with the
estimate of Woo et al. (1996) who obtained $\dot{P}_{orb}/P_{orb} =
(-5.3\pm 2.7) \times 10^{-7}$ yr$^{-1}$.  The coefficients for the
best fit quadratic function that gives the predicted times of orbital
phase zero in LMC~X-4 may be found in Table 1.

We obtain a value for the pulse period (Table 1) which is barely
shorter than that measured in 1991 with \Rosat\ ($13.50292 \pm
0.00002$ s; \cite{saf96,woo96}).  The average value of the pulse
period derivative over the intervening $\sim 7$ years is
$\dot{P}_{pulse}/P_{pulse} = (-4.9 \pm 0.2) \times 10^{-6}$ yr$^{-1}$
which is much smaller in magnitude than the derivative which we
estimate from our timing analysis of the \RXTE\ observations, i.e.,
$\dot{P}_{pulse}/P_{pulse} = (-2.14\pm 0.02) \times 10^{-3}$
yr$^{-1}$.  A graph of the history of the pulse period in Woo et
al. (1996) shows that episodes of both spin up and spin down have
occurred during the last 20 years.  Since the long-term average spin
up (or spin down) rate of LMC~X-4 is much smaller in magnitude than
the rate typically measured during an observation of a few weeks or
less in duration, the pulsar may be near its equilibrium rotation
period (see also \cite{nar85,woo96}).

\subsection{Large Flares}

It is believed that the X-ray luminosity of LMC X-4 is usually fairly
constant at a value somewhat higher than the Eddington luminosity for
a 1.4~M$_{\sun}$ neutron star, even though the X-ray flux observed at
Earth is modulated with a 30.5-day period (\cite{lan81}).  It has been
known for a long time, however, that LMC X-4 exhibits large X-ray
flares, which typically last for $\sim1000$ sec and which occur
non-periodically approximately every few days (\cite{kel83a,lev91} and
references therein).  During the flares, the 2-25 keV X-ray luminosity
increases by factors of $\sim2$ to 5, and the X-ray spectrum softens
considerably.

Flares were apparent during the \RXTE\ observations of 1998 October 14
(Julian Date $\simeq 2451100.9$, \RXTE\ observation
no. 30085-02-01-00), October 26 ($\simeq 2451112.9$, 30085-01-27-02),
and October 28 ($\simeq 2451114.1$, 30085-01-29-00).  A plot of the
count rate during a portion of the largest of these flares (that of
1998 October 28) is shown in Figure 6. The observations were
interrupted just after a peak in the flare by an Earth occultation of
the source as seen from the moving \RXTE\ spacecraft.  The top panels
of Figure 6 show the count rate of LMC X-4 in 2-sec time bins for the
8-20 keV and 2-8 keV energy bands.  Three temporal structures are
evident in the raw counting rate: the 13.5-sec pulsations, a
modulation with a characteristic timescale of $\sim150$ sec, and an
overall outburst time of $> 400$ sec.  The bottom panel of Figure 6
shows the ratio of the count rate in the 2-8 keV to that in the 8-20
keV energy channel, i.e., a ``softness'' ratio.  As was found in
previous observations of these flares, they are spectrally very soft.
Note also that the pulsations essentially disappear in the softness
plot, indicating that the softness is a monotonically increasing
function of the source luminosity and is not a function of pulse
phase.

The peaks of the X-ray counting rates during this particular flare are
up, compared with the quiescent level, by factors of 4 and 7 in the
8-20 keV and 2-8 keV bands, respectively.  After properly integrating
over the 2-25 keV X-ray band and averaging over at least one pulse
cycle, we find that the X-ray luminosity increased by a factor of
$\sim3$, which yields an absolute value for the peak luminosity of the
flare of $\sim 6 \times 10^{38}$ ergs s$^{-1}$.

We discuss physical implications of the large X-ray flares in
section 4.2.

\subsection{Pulse Profiles}

We used the best-fit orbital and pulse frequency parameters to produce
three sets of pulse profiles for each of 10 energy intervals (Figures
7-9).  The profile sets are averages for (1) the same 12 observations
used to form the template pulse profile (interval A of Fig. 3), (2)
the portion of the observation of 1998 October 28 during which a
strong flare was observed (interval B), and (3) three of the later
observations (interval C) in which the 8 - 20 keV pulse phase was
quite different from that anticipated from the 28 observations used
for the pulse timing fits.

The profiles in Figure 7 can be compared with the non-flare profiles
obtained from \Ginga\ observations in 1988 (shown in \cite{woo96}) and
1989 (shown in \cite{lev91}).  As a rough approximation, the profiles
can be thought of as consisting of an underlying sinusoid with phase
and strength that may vary with photon energy and a prominent pair of
narrow absorption dips.  In both the \Ginga\ and \RXTE\ sets of
non-flare profiles the narrow dips (at phases $\sim0.58$ and 0.82 in
Figure 7) are separated by $\Delta\phi \sim 0.25$ and the first dip is
more prominent, i.e., deeper, than the second.  In general, the dips
exhibit structure on time scales down to the time resolution of the
folded profiles, e.g., $\sim0.27$ s in Figure 7.  In the \Ginga\
profiles the underlying sinusoidal profile appears to be stronger at
low energies and shifted in phase (relative to the dips) in comparison
with the underlying sinusoid in the present \RXTE\ profiles.

Average pulse profiles during the time of the strong flare are shown in
Figure 8.  The epoch of pulse phase zero is the same as for the
profiles shown in Figures 7 and 9.  In general, the pulse
profiles during the flare are simpler than those outside of the flare
and are essentially sinusoidal at all energies studied.  Moreover,
during the flare the pulses are in phase at all energies, in contrast
with the profiles outside of the flares.  We also note that the flare
profiles are very similar to the flare pulse profiles obtained from
\Ginga\ observations in 1989 (see \cite{lev91}).

The set of pulse profiles produced from data taken late in the \RXTE\
observation of LMC~X-4, and during an interval when there was no
obvious flaring activity, is shown in Figure 9.  These profiles are
very similar to those obtained during the flare (see Figure 8) in that
they are rather sinusoidal in shape, and in phase at all energies.
However, the fractional amplitude of the modulation is much lower in
the profiles in Fig. 9 than in the flare profiles.  This is a dramatic
example of how the pulse profiles can change long ($\sim$ a day) after
the occurrence of a major flare, although it is possible that strong
flares may have occurred closer to these later observations during the
large gaps in observational coverage. It is changes in pulse shape
such as these that account for the occasional erratic behavior of the
pulse phase, as exhibited in Figure 3.

\section{DISCUSSION}

\subsection{Orbital Period Changes}

Levine et al. (1993) discuss orbital period changes in the context of
the measured orbital period derivative of SMC X-1.  They derive an
estimate for the orbital period derivative for the situation where
expansion of the companion star drives the orbital decay.  Their
discussion is pertinent when the logarithmic expansion rate of the
companion is roughly constant for at least the time necessary to
establish a pseudo-equilibrium in the difference between the rotation
frequency of the companion star and the orbital frequency.
In this case the orbital period derivative is given by (eqn. 6 of
\cite{lev93})
\begin{equation}
\frac{\dot{P}_{orb}}{P_{orb}} \simeq - \frac{\omega_{c} \, d \ln (I)/dt}
{\omega_{K}(\mu a^{2}/3I - 1)}
\end{equation}
where $\omega_{c}$ and $\omega_{K}$ are, respectively, the angular
frequencies of the companion's rotation and the orbital motion, $I$ is
the moment of inertia of the companion, $\mu$ is the system reduced
mass, and $a$ is the separation of the centers of the two stars.

Monte Carlo analysis of the binary system parameters of LMC X-4 yields
an estimated mass of the companion of 12 to 18 M$_{\sun}$ (see
\cite{jos84,nag89,lev91}).  Stellar evolution calculations for a 15
M$_{\sun}$ star (Ph. Podsiadlowski 1991, private communication) show
that the rate of increase of the companion star's moment of inertia
can be as large as $d \ln (I)/dt \sim 3 \times 10^{-7}$ yr$^{-1}$ near
the end of the main-sequence phase when the supply of hydrogen in the
star's core is nearly exhausted.  The Monte Carlo analysis of the
system parameters also indicates that the factor $(\mu a^{2}/3I - 1)$
most likely has a value in the range of 0.3 to 1.1.  This would then
yield an orbital period derivative $\dot{P}_{orb}/P_{orb} \sim - (3 -
10) \times 10^{-7}$ yr$^{-1}$ which is compatible with the measured
value $\dot{P}_{orb}/P_{orb} = (-9.8\pm 0.7) \times 10^{-7}$
yr$^{-1}$.

The radius of the companion star is estimated from the Monte Carlo
analysis to be in the range 6 - 9 R$_{\sun}$
(\cite{jos84,nag89,lev91}) which is compatible with the radius near
the end of the main-sequence phase derived by Podsiadlowski.

One problem is that the optically-observed luminosity and temperature
place the companion in a region of the H--R diagram which indicates a
mass of $\sim 25$ M$_{\sun}$, which is above the range indicated by
the Monte Carlo calculations.  We note that the calculations of
Podsiadlowski did not incorporate several effects which could be
important in the determination of the companion mass, e.g., mass loss
via a wind, effects of the evolution of the companion in a binary
system with the likelihood that substantial mass transfer has
occurred, the low metallicity of stars in the LMC, and effects of
X-ray illumination.

Our measured orbital period derivative also firmly establishes that
the orbit of LMC~X-4 is decaying, as are the orbits of SMC X-1 and Cen
X-3, on a timescale of about half a million years.  Since the
photospheric gas-pressure scale height of the massive companion star
is expected to typically be about $0.1 R_{\sun}$, then as the orbit
decays the Roche lobe will descend into the companion's atmosphere by
an additional scale height every $\sim$10,000 yr.  Thus, the mass
transfer will run away to super-Eddington rates over a relatively
short timescale.  This process will presumably end with the neutron
star spiraling into the envelope of the companion, thereby terminating
the high-mass X-ray binary phase of the evolution.  See Levine et
al. (1993) for an extensive discussion of this scenario in the context
of SMC X-1.

\subsection{Large X-Ray Flares}

The three large flares seen in the present observations are similar in
all respects to the flares of LMC~X-4 observed in detail on earlier
occasions (\cite{lev91} and references therein).  During such flares
the X-ray luminosity (2-25 keV) goes from its typical nearly steady
value of $\sim2 \times 10^{38}$ ergs s$^{-1}$ up to $\sim 10^{39}$
ergs s$^{-1}$, and the modulation factor of the pulsations increases
substantially.  As the luminosity (computed using the flux averaged
over a pulse cycle) increases by factors of 2 to 5, the X-ray spectrum
softens considerably.  While the X-ray spectrum during neither the
flaring nor quiescent modes is well fit by a blackbody, we can say
that the characteristic temperature associated with the emission
decreases by about a factor of 2 during the flaring state.  This was
shown for the instance of the flare observed on 1998 October 28 by
fitting the spectrum with the exponential form $e^{-E/kT}$.  The total
energy released in a flare is typically $\sim 10^{41}$ ergs.  Finally,
we point out that no other known accretion-powered X-ray pulsar
exhibits such flares.

From the perspective of a simple thermalization scenario, a factor of
2 to 5 increase in luminosity, coupled with a decrease in temperature
by a factor of $\sim2$, leads us to infer that the emitting area
increases by a factor of about 50!  For a nominal polar cap surface
area of $\sim3$ km$^2$, which might correspond to the accreting area
during the quiescent state, the emitting area during the flare would
have to increase to more than $\sim100$ km$^2$. This is equivalent to
the surface area of a cylindrical accretion column 25 km tall with a
diameter of 2 km, or of a cap that extends $\sim\,40\arcdeg$ outward
from the pole.

The obvious question that arises is ``what is the physical origin of
the large flaring events?''  We discuss briefly three possible origins
for the flares: (1) an instability which dumps $\sim 10^{21}$ grams of
matter stored in the accretion disk or magnetosphere, with the
attendant release of gravitational potential energy; (2) the release
of nuclear energy due to the fusion of carbon and similar elements
below the surface of the neutron star; and (3) an event involving the
release of magnetic energy.

It is commonly thought that variations in accretion rate are
responsible for the ubiquitous variability and the flaring
seen in a large fraction of X-ray binaries.  Accretion events must
therefore be considered as candidate sources of energy for the LMC X-4
flares.  It is of interest to note that the energy released by the
Type II bursts that are attributed to accretion instabilities in GRO
J1744--28 can be comparable with that released by the flares in LMC
X-4; furthermore, the GRO J1744-28 bursts may exceed the Eddington
luminosity for a canonical neutron star by factors up to $\sim100$
(\cite{jah98}).  On the other hand, the flaring in LMC X-4 is quite
distinct in detail from the variability and flaring in other X-ray
binaries, including the Type II X-ray bursts in the Rapid Burster (MXB
1730--335) and GRO J1744--28 (\cite{gue99,nis99,wds99} and references
therein).  Moreover, in a high-strength magnetic field pulsar such as
LMC X-4, the inferred increase in surface area during the flares does
not seem to have a ready physical interpretation.  For example,
emission from the sides of an accretion column might require a height
of more than a neutron star radius.  In order to confidently retain
the picture of the flares as accretion events, one would like to be
able to model accretion onto the polar caps of a neutron star at the
Eddington limit, increase the rate by a factor of $\sim5$ for
$\sim500$ sec, and compute the expected spectrum at least well enough
to demonstrate the observed softening.  To our knowledge, this has not
been done.

An energy release of $\sim 10^{41}$ ergs, within the context of a
nuclear burning scenario (\cite{bb98}) would require burning $\sim
10^{23}$ g of carbon and similar elements, with a nuclear burning
efficiency of $\sim10^{-3}$.  (Hydrogen and helium are burned almost
as quickly as they are accreted very near the surface of the neutron
star.) This scenario has the advantage that if the nuclear burning
takes place over a region of the neutron star surface that is larger
than the footprint of the accretion flow, the flare spectrum would
naturally be expected to be softer than the quiescent spectrum.  At
accretion rates of $10^{18}$ g s$^{-1}$, the mean time between
outbursts would be $\sim 10^5$ s ($\sim 1$ day), which is roughly
typical of flare recurrence times.  The large and fairly constant
$\sim50\%$ modulation factor of the pulses during the flares, compared
with only $\sim20$\% away from the flares, might be explained by
subsurface nuclear burning that takes place at only one of the two
magnetic poles during a particular event, so that substantial
modulation would be observed as the polar cap rotates in and out of
view.  However, the subflare structure observed on a timescale of
$\sim150$ s lacks a natural explanation in the context of a deep
nuclear burning scenario.  Finally, in this regard we note that Brown
\& Bildsten (1998) found difficulties explaining the large flares in
LMC~X-4 with carbon burning.  The principal difficulty is the fact
that accretion of H-rich material is thought to produce much heavier
``ashes'' than carbon, in which case not much carbon would be expected
to accumulate and be available to power the flares.

Finally, we consider the possibility that magnetic energy powers the
flares.  Magnetic energy might be suddenly released, e.g., if the
accreting material at the polar caps were somehow able to compress the
magnetic field until an instability leads to a reconnection of the
field lines.  The energy of $\sim 10^{41}$ ergs in a flare would
require that the product of magnetic field and volume be
\begin{equation}
B_{13}^2 V_{km^3} \simeq 25
\end{equation}
where $B_{13}$ is the surface magnetic field strength in units of
$10^{13}$ G, and $V_{km^3}$ is the volume in units of cubic
kilometers.  For $B_{13} = 1$, the magnetic energy contained within
$\sim 1$\% of the neutron star volume or, alternatively, of its
magnetosphere would need to be released.  If this is true, then the
field would need to be regenerated, since observations suggest that
flares have occurred every few days over the past 20 years, at least.
To our knowledge, the detailed physics of such a scenario has not yet
been explored.

A number of similarities between LMC~X-4 and soft gamma repeaters
(SGRs, \cite{kou99,hur00}) suggest a possible connection within the
context of the magnetar model (\cite{td95}).  First, even though the
large flares in LMC X-4 last much longer than the bursts from SGRs,
the energies of the flares and bursts are roughly comparable; the
flares in LMC X-4 have fluences of $\sim\,10^{41}$ ergs and SGR bursts
have fluences often in the range $10^{38}$ to $10^{41}$ ergs
(\cite{gog00}).  Second, if the neutron star in LMC~X-4 is rotating at
about the equilibrium rate estimated in the standard accretion torque
model (\cite{lpp73,raj77,ghl79,bil97}), this could imply a surface
dipole field strength of $\sim 3 \times 10^{13}$ G.  This is stronger
than the canonical $10^{12}$ G field strength thought to be typical of
accretion-powered pulsars, but is somewhat below the field strengths
thought to characterize magnetars.  However, we note that Makishima et
al. (1999) estimate a surface field strength for LMC~X-4 of $\sim\,2
\times 10^{12}$ G based on the shape of its high energy
spectrum. Third, both LMC X-4 and the persistent pulsing sources
associated with SGRs have notable variations in pulse profile shapes
(C. Kouveliotou 2000, talk given at Rossi 2000, Greenbelt, MD).
Fourth, two of the five known SGRs are nearly coincident in direction
with clusters of massive stars (\cite{fuc99,vrb00}), which suggests
that SGRs might be associated with massive stars.  Perhaps the facts
that the LMC X-4 flares are longer in duration and have softer spectra
than the bursts of soft-gamma repeaters is related to a lower magnetic
field strength and the presence of accretion in LMC~X-4.

Thus, at this point, we are uncertain as to whether the origin of the
energy in the flares is gravitational, nuclear, or magnetic.  However,
LMC X-4 is unique in its flaring behavior, and the resolution of the
origin of the flares may well provide a key to understanding some
fundamental property of the neutron star.

\acknowledgements

The authors gratefully acknowledge useful conversations with Lars
Bildsten, Andrei Gruzinov, Ed Morgan, Mike Muno, and Henk Spruit.  We
also thank Mike Muno for assistance with the spectral analysis.  This
work was supported in part by NASA Grant NAG5-8244 and by NASA
Contract NAS5-30612.

\begin{deluxetable}{lccc}
\tablewidth{0pt}
\tablecaption{Orbital and Pulse Parameters for LMC X-4\label{table1}}
\tablehead{
\colhead{Parameter} & 
\colhead{Units} &
\colhead{Value\tablenotemark{a} \ (2-8 keV)} & 
\colhead{Value\tablenotemark{a} \ (8-20 keV)}
}
\startdata
$a_{x} \sin i$\tablenotemark{b} & lt-s & $26.333 \pm 0.019$ & $26.370 \pm 0.031$ \nl
$e$        &   & $<0.003$ $(2 \sigma)$ & \nl
$T_{\pi/2}$\tablenotemark{c} & MJD\tablenotemark{d} \ TT & \ $51110.86571 \pm 0.00012$  & $51110.86600 \pm 0.00020$ \nl
$\nu_{pulse}$\tablenotemark{e,f} & Hz & $0.074060687 \pm 0.000000010$ & $0.074060697 \pm 0.000000016$ \nl
$\dot{\nu}_{pulse}$\tablenotemark{e} & Hz s$^{-1}$ & $(5.01 \pm 0.04) \times 10^{-12}$  &
			$(5.11 \pm 0.07) \times 10^{-12}$ \nl
$\ddot{\nu}_{pulse}$\tablenotemark{e} & Hz s$^{-2}$ &	$(-1.4 \pm 0.7) \times 10^{-18}$ &
			$(-1.2 \pm 1.2) \times 10^{-18}$ \nl
$a_0$\tablenotemark{g} & MJD TT & $48,137.7500\pm 0.0004$ & \nl
$a_1$\tablenotemark{g} & days &  $1.40839776\pm 0.00000026$ & \nl
$a_2$\tablenotemark{g} & days &  $(-2.65\pm 0.19) \times 10^{-9}$ & \nl
$\dot{P}_{orb}/P_{orb}$ & yr$^{-1}$ & $(-9.8\pm 0.7) \times 10^{-7}$ & \nl
\enddata
\tablenotetext{a}{The errors ($1 \sigma$) have been obtained from a
   least-squares analysis in which the $1 \sigma$ errors of the
   observed pulse arrival times were estimated by the root mean square
   deviation of the measurements from the best fit model.  This may
   underestimate the effects of certain types of systematic errors.}
\tablenotetext{b}{$a_x\sin{i}$ weighted average $= 26.343 \pm 0.016$ lt-s}
\tablenotetext{c}{$T_{\pi/2}$ average = MJD $51110.86579 \pm 0.00010$
   (TT) (but see text)}
\tablenotetext{d}{Modified Julian Date $=$ Julian Date $- \ 2,400,000.5$}
\tablenotetext{e}{For the epoch MJD 51235.0}
\tablenotetext{f}{$P_{pulse} = 13.502440 \pm 0.000002$ s (2-8 keV) or $13.502439
   \pm 0.000003$ s (8-20 keV)}
\tablenotetext{g}{The predicted times of orbital phase zero are given by: 
   $t_N = a_0 + a_1 N + a_2 N^2$.}
\end{deluxetable}

\begin{figure}
\figurenum{1}
\epsscale{1.0}
\plotone{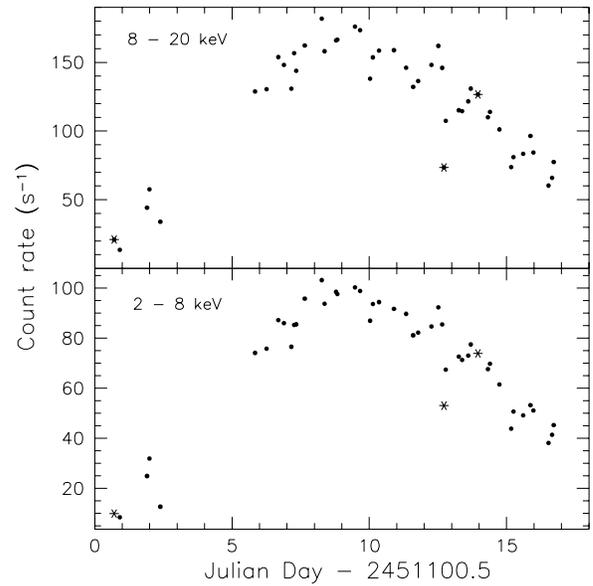}
\caption{Average background-subtracted count rates in 2 energy bands
for each of the 47 observations of LMC X-4 obtained in 1998 October
with the PCA on \RXTE\,. The observations in which strong flares
occurred are denoted by ``$\ast$''; however, only the data taken
during non-flaring intervals were used in computing the average rates
shown for these observations.  Time intervals during which LMC X-4 was
in eclipse were also excluded in the calculation of the average rates.
In the few instances when fewer than 5 PCUs were in operation, the
rates have been normalized to a 5 PCU basis. Estimated errors in the
average rates due to counting statistics are negligible.  Julian Date
2,451,100.5 corresponds to 1998 October 14 $0^{\rm h}$ TT. \label{fig1}}
\end{figure}

\begin{figure}
\figurenum{2}
\epsscale{1.0}
\plotone{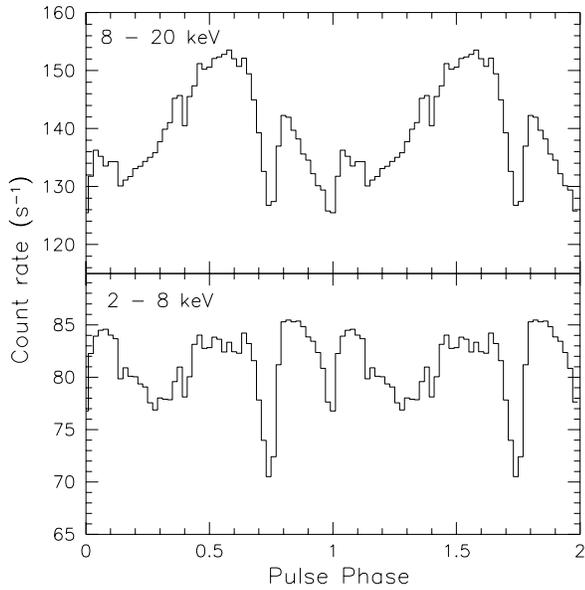}
\caption{Average pulse profiles for 2 energy bands for 12 observations
of LMC X-4 (see interval A in Fig. 3).  These profiles were used as
the templates for our pulse timing analysis.  The $1 \sigma$
uncertainties from counting statistics are approximately 0.5 counts
s$^{-1}$. \label{fig2}}
\end{figure}

\begin{figure}
\figurenum{3}
\epsscale{1.0}
\plotone{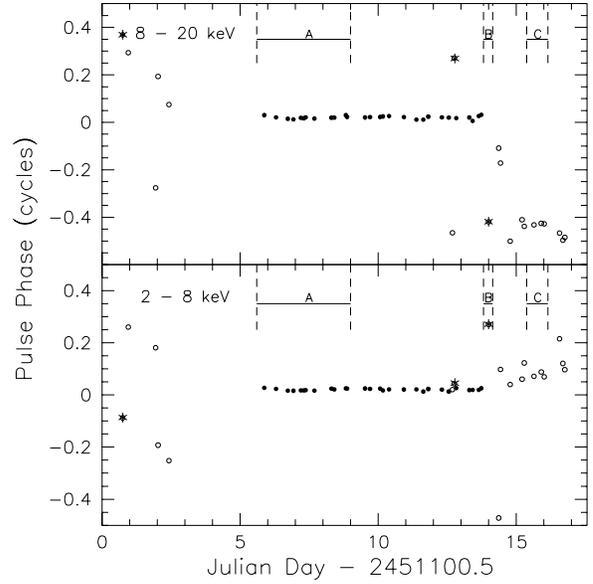}
\caption{Results of our pulse timing analysis for 2 energy bands.
Each symbol gives the phase of the pulse profile for one of the 47
observations of LMC X-4 relative to the phase of a model pulsar.  The
solid circles indicate measurements that we use in the least-squares
fit to obtain orbital and spin parameters.  The open circles are
measurements that are not used in the fit because the timing results
here indicate significant inconsistencies from the results expected
from a pulsar with an invariant pulse profile.  The observations in
which strong flares occurred are denoted by ``$\ast$''; however, the
data taken during the flares were not included in the pulse profiles
made for the timing analysis.  Time interval A was used for the
construction of the template pulse profiles (Fig. 2).  Time intervals
A, B, and C were used to make the energy-dependent profiles shown in
Figures 7-9, respectively. \label{fig3}}
\end{figure}

\begin{figure}
\figurenum{4}
\epsscale{1.0}
\plotone{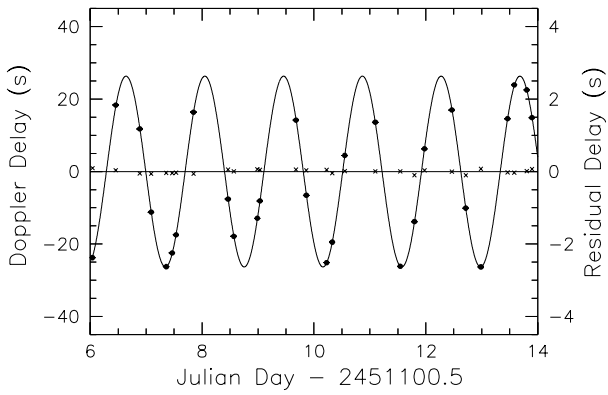}
\caption{Inferred pulse arrival time delays for 28 observations of LMC
X-4 (see Fig, 3 and text).  The solid circles and sine wave fit show
the delays with respect to a model pulsar at the center of mass of the
binary system (scale on left), while the crosses show the measured
time delays relative to a pulsar in the best fit orbit (scale on
right). \label{fig4}}
\end{figure}

\begin{figure}
\figurenum{5}
\epsscale{1.0}
\plotone{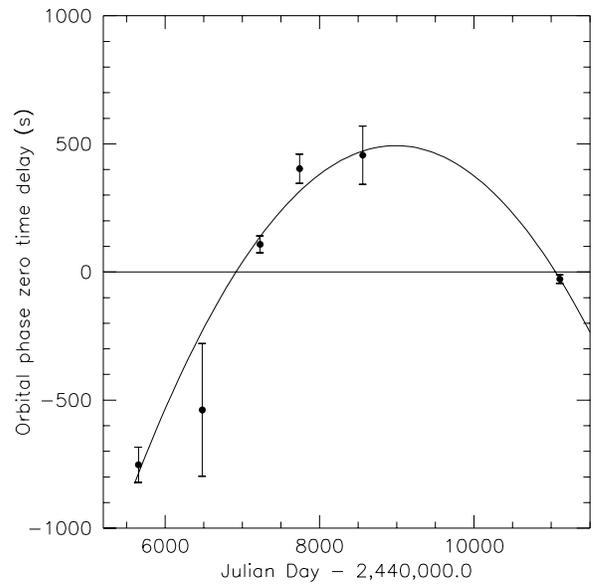}
\caption{Epochs of orbital phase zero of LMC X-4 are shown relative
to the epochs expected according to the best-fit constant orbital
period of 1.40839457 days. The measurements are based on \EXOSAT\,,
\Ginga\,, and \Rosat\ data (Dennerl 1991; Levine et al. 1991; Woo et
al. 1996) and the present work.  The curve shows the best-fit
quadratic function.  The error bars shown indicate $\pm 1 \sigma$
uncertainties. \label{fig5}}
\end{figure}

\begin{figure}
\figurenum{6}
\epsscale{1.0}
\plotone{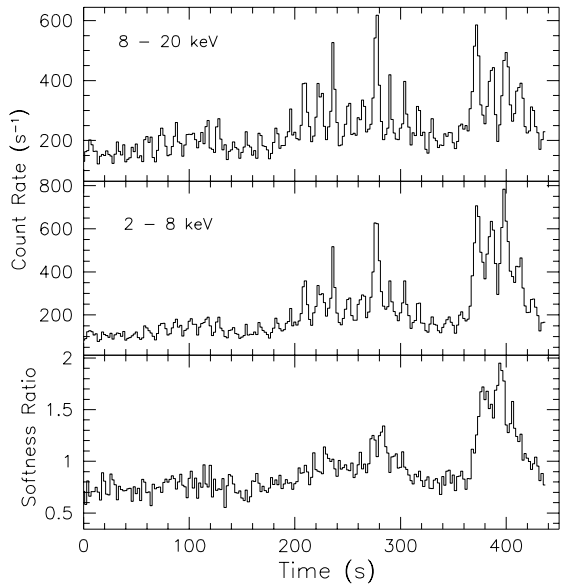}
\caption{(top and middle) Background-subtracted count rates for 2
energy bands during a portion of the observation of 1998 October 28
(\RXTE\ observation no. 30085-01-29-00) when a strong flare was in
progress. See Section 3.2 for a detailed description of the flare.
The observation was terminated by Earth occultation of the source at
about 430 s while the flare was in progress.  The 13.5 s pulsations
are highly evident in both energy bands.  (bottom) The softness ratio
defined by the number of counts in the 2--8 keV band divided by the
number in the 8--20 keV band.  The rates and the softness ratio are
averaged in 2 s time bins. The 13.5 s pulsations are not evident in
the softness ratio plot. \label{fig6}}
\end{figure}

\begin{figure}
\figurenum{7}
\epsscale{1.0}
\plotone{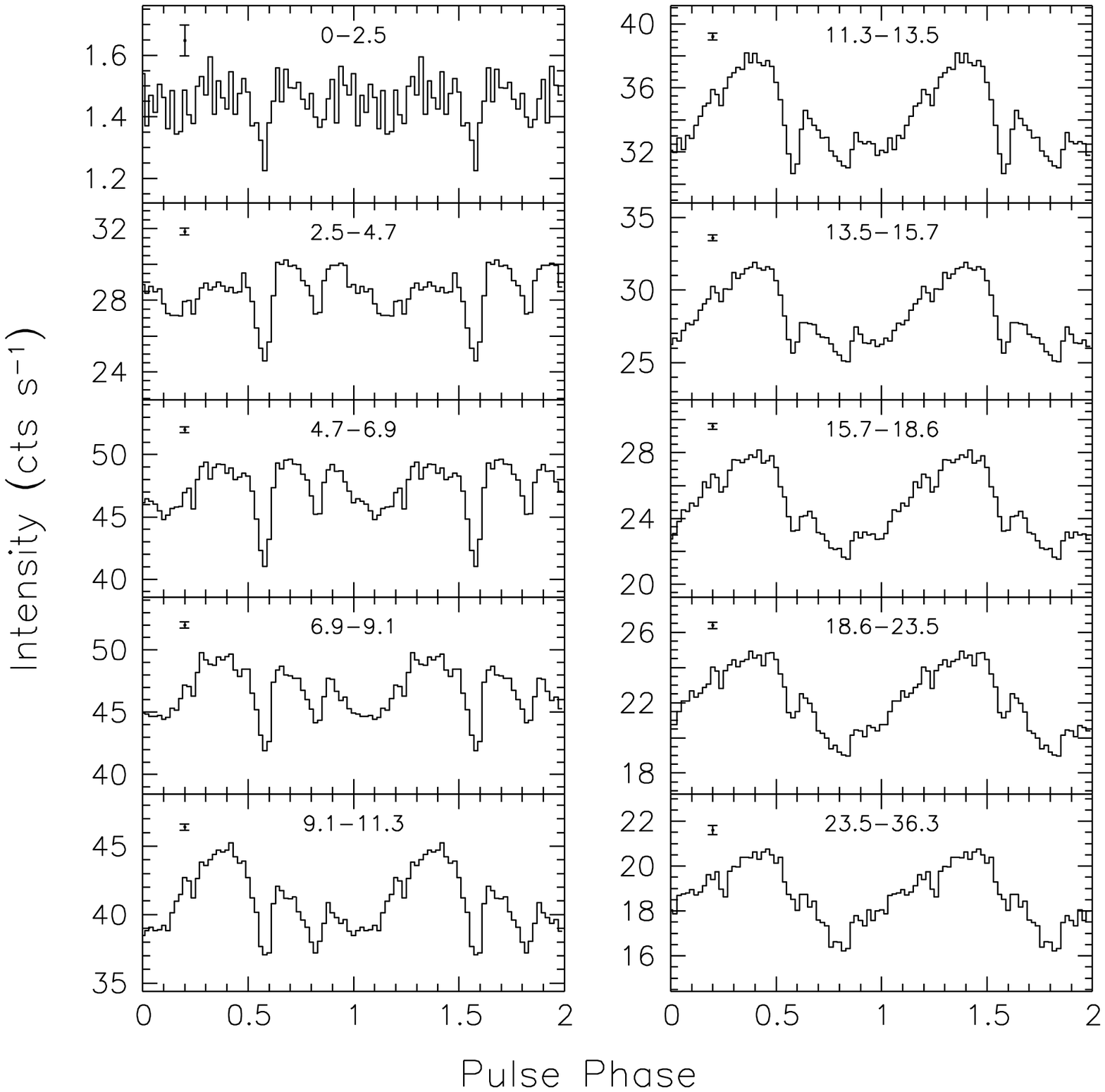}
\caption{Average background-subtracted pulse profiles for 10 energy
bands for the 12 observations in time interval A (see Fig. 3).  Each
profile is plotted twice for clarity.  A typical $\pm 1 \sigma$ error
and the energy range in keV are shown for each profile. \label{fig7}}
\end{figure}

\begin{figure}
\figurenum{8}
\epsscale{1.0}
\plotone{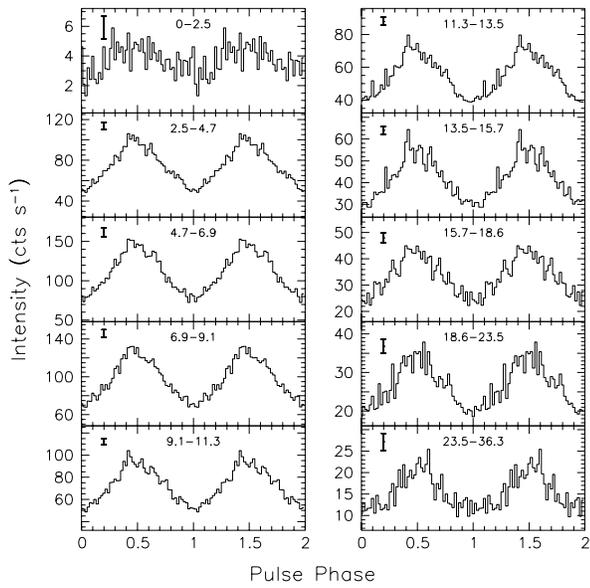}
\caption{Average background-subtracted pulse profiles for 10 energy
bands for the observation of the flare which occurred on 1998 October
28 (during interval B of Fig. 3).  The profiles were constructed using
data from the entire time interval plotted in Fig. 6.  Each profile is
plotted twice for clarity.  A typical $\pm 1 \sigma$ error and the
energy range in keV are shown for each profile. \label{fig8}}
\end{figure}

\begin{figure}
\figurenum{9}
\epsscale{1.0}
\plotone{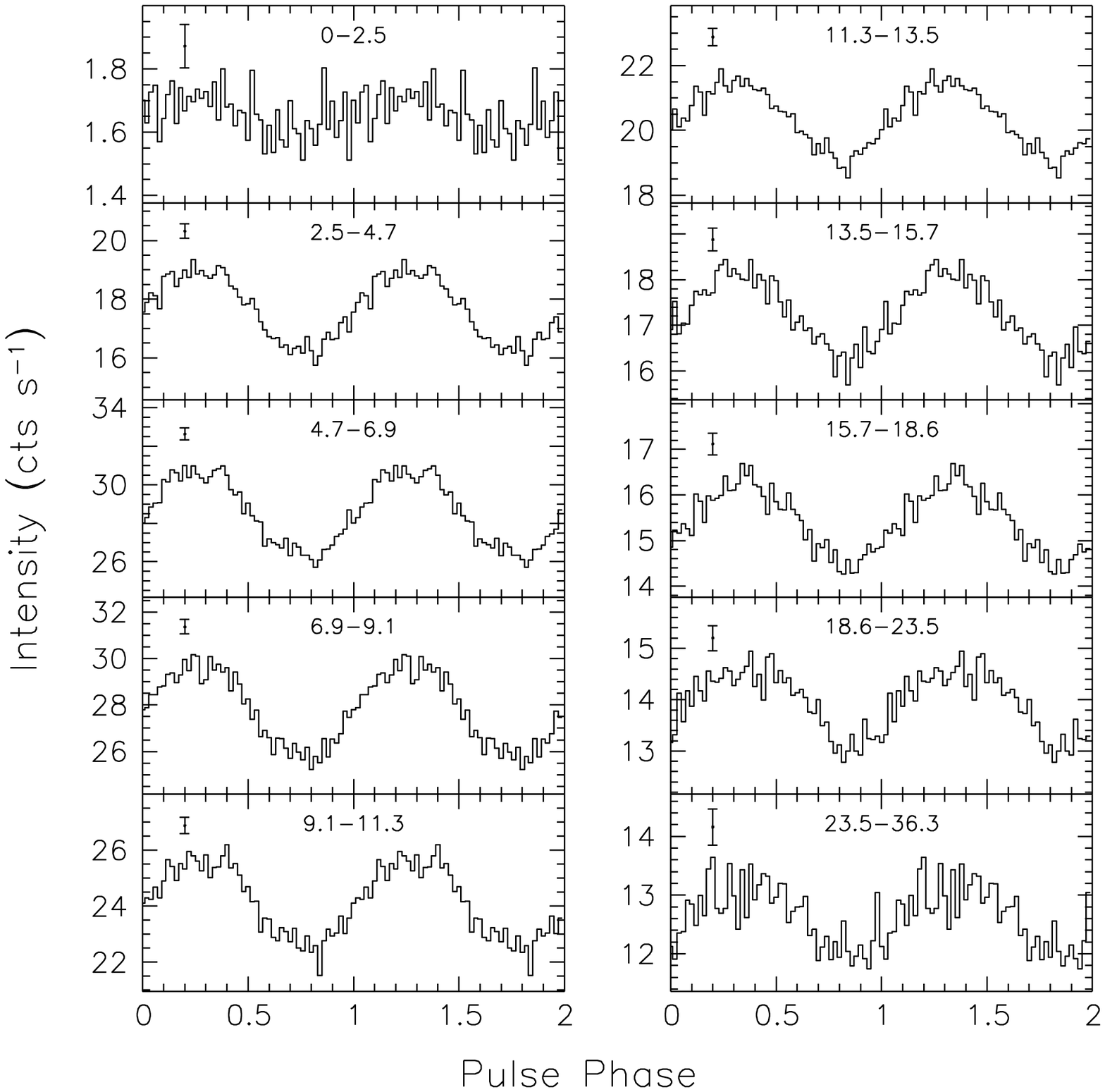}
\caption{Average background-subtracted pulse profiles for 10 energy
bands for the 3 observations in time interval C (see Fig. 3).  Each
profile is plotted twice for clarity.  A typical $\pm 1 \sigma$ error
and the energy range in keV are shown for each profile. \label{fig9}}
\end{figure}

\end{document}